\documentclass[10pt,a4paper, titlepage]{article}
\usepackage[utf8]{inputenc}
\usepackage[font=small,labelfont=bf]{caption}
\usepackage{graphicx}
\usepackage{chngpage}
\usepackage{a4wide}
\title{Ensuring anonymity in survey panels}
\author{Kees Mulder}

\begin{document}



\begin{titlepage}
\begin{center}
\textsc{\Large }\\[0.5cm]
{ \huge \bfseries Ensuring anonymity in survey panel research \\[0.4cm] }
\begin{minipage}{0.9\textwidth}
\begin{flushleft} 
\emph{Author:}\\
Kees Mulder
\end{flushleft}
\end{minipage}
\\[1.0cm]
\begin{minipage}{0.9\textwidth}
\begin{abstract}
In survey panel research, anonymity of the participants is of great importance, as it must be ensured to prevent negative effects of participation as well as to maintain trust that the sensitive data that respondents provide is handled with care. Measures have been developed in the disclosure control literature to provide estimates for re-identification risk in a dataset. Applying them to survey research is not straightforward as most methods that have been developed do handle missing data properly. Here, a new method is applied that closely matches a realistic scenario of an attempt of re-identification for survey research. This method can be applied to assess re-identification risk in survey research.  
\end{abstract}
\end{minipage}
\vfill
{\normalsize \today}

\end{center}
\end{titlepage}
\newpage

\section{Introduction}

In social research, a growing demand exists for (semi-)public use of survey panel microdata, that is, raw data containing separate information for each respondent. There are many benefits to using large scale survey panels rather than separate data-collection for each study. Researchers are saved the logistic operations and can focus fully on their research, large samples can be maintained so that statistical power is usually high, longitudinal data can be included easily, and data from many studies can be combined, as they were collected on the same sample. Several large survey panel organizations exist providing anonymized microdata to academic researchers and businesses alike. 

In survey panel research, anonymity is generally regulated by law. In the Netherlands, these laws clearly define the situations in which personally identified information can be disclosed. In addition to legal concerns, guaranteeing anonymity is beneficial for the scientific community as it instills trust that the researchers will handle private and sensitive data with care, which may prevent a decrease in response rates and maintain honest reporting of sensitive data (e.g. less social desirability) \cite{cho1999privacy}. Thus, anonymity of the respondents is of utmost importance. Before survey microdata can be disclosed, the agency maintaining this data must protect it against statistical disclosure \cite{willenborg2001elements}. Because of this, the recruitment of respondents and collection of data in survey panels is often separated from the analysis of the data, and data is usually anonymized before it is released to researchers for analysis.

Anonymity is achieved when (sensitive) data that were released can not be linked to a specific respondent, that is, the respondent can not be \textit{re-identified}. When personal identifiers are unavailable, re-identification efforts focus on uniquely linking the non-censored information in the dataset to another source of information on a respondent. This includes not only directly matching information between the data and a person, but also linking the released dataset to additional datasets from which personally identifying information \textit{can} be obtained. Therefore, it is inadequate (yet commonplace) to simply censor heavily identifying information, such as names, dates of birth, and geographical information. As more and more data is collected on a group respondents, it becomes increasingly likely that respondents can be re-identified on the basis of their response pattern on the non-censored variables. Investigation is then necessary into the re-identification risk of respondents by using more advanced methods.

In the re-identification and disclosure control literature, datasets used are most often register data, medical data or automatically collected data, for example from users of websites 
. Several properties distinguish such data from data obtained in survey panels. Survey panel data typically contains more opinion variables, more often includes longitudinal data, often contains much more missings or questionnaires completed by only a fraction of the sample and may be accessed by a larger amount of people, especially if access is granted to the scientific community at large. In addition, some survey panels also employ random sampling, whereas many other large sets of microdata are based on convenience samples or population registers.

In this article, knowledge from the disclosure control literature will be applied to the field of survey panel research. The adapted methods will tested by assessing the risk of re-identification for a specific survey panel, the LISS (Longitudinal Internet Studies for the Social sciences) panel administered by CentERdata (Tilburg University, The Netherlands). First, section \ref{theory} provides an overview of the theoretical framework as described in the disclosure control literature. Section \ref{methodsresults} describes how the current disclosure control methodology can be adapted for application in survey research. In section \ref{conclusion}, the implications of these results for survey research are discussed.

\section{Theoretical framework \label{theory}}

\subsection{Definitions}

First, \textit{re-identification} must be defined. In this article, a successful re-identification is taken to mean any form of connection that is made between a specific respondent in the disclosed dataset and information from any source outside of the disclosed dataset, in such a way that it becomes known, with reasonable certainty, that a person in the population is the respondent in the disclosed dataset. 

In the disclosure control literature, the \textit{disclosed dataset} is the anonymized (that is, lacking personal identifying information) dataset to which the researcher has access. Often, it may contain sensitive data, so re-identification of the respondents in this dataset should be prevented. In the present study, this refers to the data obtained through the LISS panel.

As true identifiers will not be available in anonymized datasets, re\--identification efforts must resort to a set of variables that can more or less uniquely identify respondents. \cite{daleniusqid1986}  termed this the \textit{quasi-identifier}: A variable or set of variables in the dataset that can be used to identify an individual probabilistically \cite{emamlongi2011}. This has also been referred to as the key variables. Several variables are typically combined into a response pattern, which can then be used to identify persons. Common variables used in quasi-identifiers are gender, age or postal code, but many other variables have been used, going as far as movie preference \cite{netflix}.
 
A \textit{linking dataset}, then, is a dataset containing a quasi-identifier that is also available in the disclosed dataset, as well as additional variables of interest, such as names or other personally identifying information. Using this dataset, we may try to re-identify respondents. A common example of a linking dataset is register data.

An \textit{anonymity set} is a set of respondents who have the same response pattern on the quasi-identifier. An anonymity set of size 1 means the respondent is \textit{unique} on the quasi-identifier. The size of the anonymity set is generally referred to as \textit{k}.

\subsection{Re-identification Risk}

Measuring re-identification risk is central to disclosure control, as it is essential to determining whether further anonymization is necessary as well as analyzing which measures of ensuring anonymity are effective.

In the broadest sense, the re-identification risk, which is the probability that a successful re-identification takes place ($Pr(identification)$), depends on both the risk that an adversary attempts re-identification on a disclosed dataset ($Pr(attempt)$), as well as the risk that this attempt is successful, which is  $Pr(identification | attempt)$. In the disclosure control literature, by far the most attention is given to the latter of the two: given that an attempt is made on the disclosed dataset, the likelihood that an adversary succeeds is discussed. The main cause of this is that the probability that someone will attempt re-identification is difficult to estimate, because it is influenced by a large amount of uncertain factors, while the possibilities of identification given an attempt are much more straightforward to describe, formalize and test. In addition, it is assumed that if identification given an attempt is very unlikely, then the probability that an attempt takes place is also low, as the attempt will most likely not be fruitful. 

\cite{elliot1999scenarios} stress that it is important to not only study the probability of identification if an attempt is made $Pr(identification|attempt)$, but to also to discuss the goals and motivations of a possible adversary. Then, it is possible to evaluate $Pr(attempt)$, the probability that an adversary will attempt re-identification. With this, we can calculate the probability we are most interested in: the probability that a \textit{successful} re-identification attempt is made on the disclosed data ($Pr(identification)$). For a thorough discussion of the possible scenarios of attack and the risk that an adversary attempts re-identification, see \cite{elliot1999scenarios}.

Applying this to survey panels, a few remarks can be made on the likelihood of re-identification attempts. Firstly, survey panels often allow access to their microdata to various institutions, businesses or, most often, the scientific community at large. Because of this, the amount of users with legitimate access is already fairly large, and the userbase may be even larger when accounting for the possibility that the survey panel data could be acquired through illegitimate means, such as account sharing, (accidental) sharing of data, or sale. Secondly, survey panel data itself tends to contain a very large amount of information on the respondents, often including sensitive data, so that it is likely that for many possible adversaries at least some part of the data is useful. Shortly put, for survey panels in general, the probability that re-identification is attempted is estimated to be fairly high because many people have access to the data, the data is likely useful and many sensitive items are available. 

The next step is to investigate the probability that an attempt will be successful, which is $Pr(identification|attempt)$. Uniqueness is often taken as a measure of this re-identification risk in the disclosure control literature \cite{emamlongi2011}. If a respondent is not unique in a dataset on the quasi-identifier that was used, so that there are multiple respondents with that same response pattern, then it is not possible to determine which set of responses on the rest of the data in the disclosed dataset belongs to the respondent we are trying to identify. If one would imagine a multitude of potential linking datasets, only lack of uniqueness in the disclosed dataset can fully protect against all possible linking datasets. 

\subsection{Sample and population uniqueness \label{uniquenesstypes}}

Two types of uniqueness can be distinguished: sample and population uniqueness. Which of them should be used to determine the re-identification risk depends the information and goals we ascribe to our adversary. 

\cite{emam2008} discuss this matter. Assume that an adversary has access to the disclosed dataset as well as to a linking dataset, including identifying information. Say the two datasets share a group of quasi-identifiers, and that their sampling is independent. Then, if the adversary would find a response pattern on the quasi-identifiers that is unique in both datasets, the adversary would still not be certain that this person was the same person, so the adversary would not be able to re-identify this person with certainty. 

The reason for this is that in the population the response pattern under investigation may not be unique: the two datasets may both have sampled a different person from the population, each having the same response pattern on the quasi-identifier. In this case, the re-identification risk is only concerned with population uniqueness: a person can only be uniquely identified between the two datasets if this person is unique on the quasi-identifiers in the population (and, as such, is also unique in the sample). As this hampers re-identification efforts, this is beneficial for our attempts of ensuring anonymity. 

However, this is not always viable. If the adversary is looking for a specific person in the disclosed dataset, knowing or assuming that person is in the sample, anonymity is concerned with sample uniqueness as well, which is more difficult to prevent than population uniqueness. In this situation, uniqueness should be prevented for any respondent in the disclosed dataset. \cite{emam2008} give three possibilities for when an adversary might be able to seek out specific persons in this way: 

\begin{enumerate}
\item The disclosed dataset  contains the whole population or has a very large sampling fraction. 
\item It can be easily determined who is in the disclosed dataset. This would be the case, for example, if an adversary has bank statements from the population, where he could check who had received money from participation in a survey panel.
\item Individuals reveal themselves as participants of the survey panel. For example, a friend of someone who works with a survey panel may tell them that they have been enrolled in the panel, or a respondent may state online that they partake in the survey panel. Then, with access to the disclosed dataset and information that the adversary already has about the respondent, the individual may be re-identified.
\end{enumerate}

The first situation is unlikely to occur in survey research, as it would most likely require access to governmental or census data with identifiers. It may be possible for some researchers, while most will not be given access to these datasets. It is difficult to foresee how common access to such datasets is, and how many organizations maintain such a dataset.

The second situation is also very unlikely for survey panels, as most employ random sampling techniques and virtually all do not publicly disclose the identity of the members of the panel or information from which this may be derived.  

The third, however, could occur in survey panels, yet is very hard to guard against. Because the adversary may have a large amount of information on the participant, many more variables could potentially be used in quasi-identifiers, which may make it easier to uniquely identify respondents. This situation is likely to be fairly rare, however, and may be mediated by professionalism from a researcher who has access to the disclosed dataset that comes across self-disclosure. Additionally, part of the blame could fall onto the participant revealing their participation in the survey panel, although it is unlikely that respondents will often consider the full consequence of revealing their participation. Nevertheless, this is a situation that must be prevented whenever possible. 

If any of these three situations occurs, the anonymization procedure must also concern itself with sample uniqueness: it must be prevented that persons exist in our disclosed dataset with a unique response pattern on a quasi-identifier. If sample uniqueness is prevented, even someone who knows a person to be in the disclosed dataset or has population records cannot uniquely identify them in the disclosed dataset, because their response pattern is not unique.

\subsection{Estimation of population uniqueness}

Often, there is no population data available on the quasi-identifiers, so population uniqueness must be estimated. Procedures for estimating population uniqueness from sample uniqueness are found in the disclosure control literature \cite{Zayatz91estimation, chen1998estimation}. \cite{skinner2002measure} expand on this by proposing a new measure of disclosure risk, $\theta$, which can be interpreted as $Pr(correct~match|unique~match)$, the probability that a found unique match is correct. The scenario under which this method operates is the situation where an adversary has found a match between a respondent chosen from the population and the disclosed dataset, but is not certain that this match is correct. In particular, this method takes into account the possibility of a false positive: a match that seems unique, but in fact are two distinct persons in either dataset that happen to have the same response pattern. 

In their paper, \cite{skinner2002measure}, first show a simulation method to estimate $\theta$, the probability that a match is correct, and then continue to provide a simple way to estimate $\theta$ as 

\begin{equation}
\hat{\theta} = \frac{n_1\pi}{n_1\pi + 2(1-\pi)n_2}
\end{equation}

where $n_1$ denotes the amount of unique response patterns, $n_2$ denotes the amount of pairs (sets of two persons with the same response pattern, so that $k=2$), and $\pi$ indicates the probability that a respondent is sampled in the disclosed dataset. Assuming that respondents are sampled from the population with an equal probability, $\pi$ is taken to be $n/N$, where $n$ is the amount of respondents in the disclosed dataset, and $N$ is the size of the population. 

\cite{skinner2002measure} suggest that this method has two advantages over estimating probabilities of population uniqueness: Firstly, this measure of disclosure risk is more realistic than estimating population uniqueness, because it was developed from the viewpoint of real-life scenarios, taking into account the possibility of incorrect matches, and secondly it allows consistent inference to be made without strong modeling assumptions. 

One disadvantage of this measure in practice is that as low frequencies for $k=1$ and $k=2$ occur, estimates often become 0 or 1, neither of which is realistic. 

\subsection{Re-identification approaches \label{approaches}}

Two main approaches to attempt re-identification can be distinguished.

The first approach uses a linking dataset to attempt to identify as many respondents as possible. An adversary here does not care who specifically is identified, but is solely interested in obtaining as much (sensitive) data as possible. In this case, we are concerned with population uniqueness, as described above. This method is analogous to the \textit{journalist scenario} in \cite{emam2008}. In practice, this approach is realized by dataset linking.

The second approach is to identify specific individuals for whom information is available. From the information available on the respondent a quasi-identifier may be constructed, which can then be used to potentially find a unique match in the disclosed dataset. This method requires one of the three criteria as described in section \ref{uniquenesstypes}. As such, in order to prevent this type of attack, sample uniqueness must be prevented. This method is analogous to the \textit{prosecutor scenario} in \cite{emam2008}. 


This article will focus on the first approach.

\subsection{Enforcing anonymity}

As uniqueness arises, decisions must be made to ensure the anonymity of respondents. A common simple criterion for ensuring lack of uniqueness is \textit{k-anonymity} \cite{samaratikanon1998, sweeneykanon2002}.

To satisfy $k$-anonymity, each response pattern on the quasi-identifier that occurs in the data must occur at least $k$ times. Then, if an adversary were to try to identify a person on a specific realization of the quasi-identifier, he would need to choose between at least $k$ different options, reducing the probability of correct re-identification to a maximum of $1/k$. 

The practice of $k$-anonymity has also been extended with the concept of $l$-diversity, where also the possibility is taken into account that the $k$ respondents that cannot be distinguished on the basis of the quasi-identifier have the same value on the sensitive variables of interest \cite{ldiv2007}. Then, the adversary still has certainty regarding this value for the respondent he is interested in. This is less of an issue in survey panels, as there are often so many variables that may be viewed as sensitive that a unique pattern on th at part of the data is highly probable.

Common methods of enforcing anonymity include removal of direct identifiers (ie. names, addresses), suppression of values that can be used in a quasi-identifier, setting thresholds for geographic information (for example, by not releasing geographic information on areas with a population smaller than 100,000), swapping a percentage of the values in the quasi-identifier, rounding continuous variables, or  aggregating extreme or rare values by recoding into more general categories, either by aggregating all values, or choosing a cut-off point from which all more extreme values are aggregated into a specified value \cite{hawala2003microdata}. In addition, noise addition algorithms have been developed as a method of enforcing anonymity \cite{brand2002microdata, fuller1993masking}. Methods have even been developed for replacing the disclosed dataset entirely with synthetic data, by use of the framework of multiple imputation \cite{rubin1993statistical, raghunathan2003multiple}. For further discussion of methods for enforcing anonymity, see \cite{winkler2004masking}. For an overview of which methods are used in practice, see \cite{felso2001disclosure} and \cite{skinner2009statistical}.

\section{Re-Identification in the LISS panel\label{methodsresults}}

The survey panel used in this study, the LISS panel, is a representative sample of Dutch individuals who participate in monthly Internet surveys. The panel is based on a true probability sample of households drawn from the population register. Households that could not otherwise participate are provided with a computer and internet connection. A longitudinal survey is fielded in the panel every year, covering a large variety of domains including work, education, income, housing, time use, political views, values and personality. 
The panel maintains roughly 8000 respondents at any time, with additional recruitment waves performed as the current panel size dropped due to drop-out. Participants are sampled by household. The response rate is relatively high 
 through use of various measures \cite{scherpenzeel2011lisswasbuilt, scherpenzeel2012recruiting}. In addition to the background variables, roughly 90 different questionnaires have been deployed so far to the sample. 

Background variables were obtained directly for the participants that were panel members, and by proxy for the members of their household that were not part of the sample (ie. this information was filled in for them by the active panel members). Background variable surveys were administered once a month, in order to track changes. As \cite{emamlongi2011} show, using longitudinal data to obtain the moment of change for some of these variables could greatly increase uniqueness, which may pose a large risk for anonymity. For example, the longitudinal nature of the disclosed dataset may be used to obtain the moment of marriage instead of simply the marital status of a respondent, provided the respondent was in the survey panel at the moment of marriage. Moment of marriage would clearly provide a more unique pattern than current marital status, and may thus put the anonymity in the disclosed dataset at risk if, for example, a marriage register was used as the linking dataset. 

\subsection{Dataset Preparation}

In order to obtain a dataset that was fit for analysis, some preparation was necessary. The LISS panel has maintained roughly 8000 panel members for several years, where naturally some attrition occured, as well as additional waves of recruitment. Information on all respondents that took part in the panel over the years was combined, resulting in a dataset of 17924 respondents. Then, persons for which only background information was obtained (who did not directly take part as a panel member but for whom information was provided by proxy) were discarded from the working dataset. As these persons did not provide more information than the basic background information, they are considered to not truly be part of the sample, and moreover successful re-identification on these respondents would only allow an adversary access to their background variables, which have low utility. 

In practice, this was ensured by removing every person that lacked either information in the background variables or lacked information in at least one other panel study. 6927 respondents only had data in at least one of the background variable waves, and no other panel studies, and were removed. After removing them, the working dataset consisted of 10997 respondents, which includes current and past members of the panel.

Each wave of background variables by itself contained many missing respondents, because they may not have been a respondent at the time of administration of the monthly questionnaire. In order to deal with this, the waves of background variables were merged into a single file, containing the most recent value that was obtained for each household member in the sample. This provided a dataset that had very few missings. The background variables gender, year of birth, position within the household, number of living-at-home children, civil status, domestic situation, primary occupation and education were entirely free of missing values.  Table \ref{tab:backgroundNA} provides an overview of the remaining background variables, and the amount of missings for each after merging.

\begin{table}[!htbp] \centering 
\begin{small}

\begin{tabular}{@{\extracolsep{5pt}} lcc} 
\\[-1.8ex]\hline 
\hline \\[-1.8ex] 
&&Proportion\\
Variable & Missings &  missing \\ 
\hline \\[-1.8ex] 
Degree of urbanization & 8 & 0.001 \\ 
Net monthly income & 1783 & 0.162 \\ 
Net monthly income in categories  & 406 & 0.037 \\  
Ethnic country of origin in categories & 2264 & 0.206 \\ 
\hline \\[-1.8ex] 
\normalsize \vspace{-0.7cm}
\end{tabular} 

  \caption{Overview of missing data the merged dataset containing background variables, where total $n = 10997$.   \label{tab:backgroundNA} } 
  \end{small}
\end{table} 

\vspace{15pt}
\noindent\textit{Month of birth}

In the background variables, both year of birth and age (derived from the date of birth at the time of filling in a questionnaire) are given for each respondent, but day and month of birth are suppressed. However, as this questionnaire was administered each month, month of birth could be estimated from the moment that age changed in the monthly waves of background variables. Respondents that had two different months in which their age had changed were pinned down to a specific month that had to contain their birthday (the first of the two months in which their age had changed). If the respondent always had all their age changes in a specific month, their month of birth may be that month, or the month before it. 

\begin{itemize}
\item[] \textbf{Example:} 
\textit{
\item Respondent A has changed age in March 2009, and March 2010, and then dropped out of the panel. His birthday may have been earlier in March, or in April after he filled out the monthly survey for April. The estimated month of birth is April/March.
\item Respondent B has changed age in June 2008, and July 2009, and June 2010, so changes have occured in both June and July. If his birthday was in May, his age change would have been found in the monthly questionnaire for June 2009, but it did not until a month later. If it was in July, it would not have changed in the questionnaire for June already, which it did. Therefore, the estimated month of birth is June.
\item Respondent C has only been a member for 7 months, and no age changes were found in that time. No month of birth is estimated, and this value is treated as missing.}
\end{itemize}

After this procedure, 8845 respondents (80.4\%) had two possibilities as an estimated month of birth, 1979 respondents (18\%) had a single estimated month of birth, and finally, 173 (1.6\%) of respondents did not have an estimate, most likely because their membership period of the panel did not include their birthday.

\subsection{Dataset Linking}

In this section, re-identification risk from dataset identification will be assessed for longitudinal survey panels. In dataset identification, respondents in a linking dataset are matched to respondents in the disclosed dataset by use of a quasi-identifier. To assess the likelihood of success of this scenario, the variables in the quasi-identifier must be known. For that, however, it must be determined which linking dataset will be used.

Many datasets have been used as linking datasets in the past, and it is difficult to determine beforehand which is available to an adversary. For example, the linking dataset could be data that companies have collected on their customers, data from Statistics Netherlands, data from public registers, or many other sources. Thus, instead of attempting to assess the risk for each imaginable linking dataset, re-identification risk may be studied for several expected or common quasi-identifiers, which correspond to as yet unknown linking datasets \cite{kootdutch2010}.

\subsubsection*{Methods}

As was shown before in Table \ref{tab:backgroundNA}, for some variables in the data, a sizable fraction of missing data was found. In addition, survey panels, including the LISS panel, often administer questionnaires to a subsample of the panel members, causing a number of observations to be missing for a (large) proportion of respondents. In order to assess the result of the availability of these variables for re-identification, the issue of missing data must be addressed. 

In the statistical disclosure control literature, missing data is often either not present in the disclosed data, or all respondents that have at least one missing value in their quasi-identifier will be discarded \cite{iyengar2002transforming, emam2008, yancey2002disclosure, bender_survey_2001}. For a longitudinal survey panel such as the LISS panel, this approach is generally not feasible. Occurrence of missing data is virtually inevitable in large-scale survey panels such as these, and their occurrence is common enough that simply removing respondents that have missing values on the quasi-identifier may mean losing too large of a fraction of the dataset. 

Several possible approaches exist towards dealing with missing data. If the missingness is interpreted simply as a value in the response pattern, values for the parameters we are interested in may be obtained. However, in that case it has been assumed that an adversary will know that a respondent will be missing on this attribute in the disclosed dataset, which is an unlikely assumption. 

Model-based imputation techniques may also be considered to be used to deal with missing data in a re-identification scenario \cite{elemam2009evaluating}, although little work has employed this technique in the literature. It would make exact matches impossible, which invalidates most methods of matching. 

A more reasonable approach is to remove respondents from the dataset if they have at least one missing value on any of the variables in the quasi-identifier, as in listwise deletion. Another approach is to attempt to find all possible matches for each of the respondents, where missing values are treated as a match for all other values. Both of these approaches will be discussed below. 

\vspace{15pt}
\noindent \textit{Listwise deletion}

When performing listwise deletion, the discarded respondents are essentially treated as if they are not part of the sample. Previous knowledge on which respondents may or may not be in the dataset is then no longer useful. 

As such, this option is reasonable for assessing re-identification risk in the scenario where an adversary has a linking dataset with a very large sampling fraction (e.g. close to the population), because in that situation, the adversary does not have information on who might be in the disclosed dataset (see section \ref{uniquenesstypes}). 

After listwise deletion, the process of assessing uniqueness is as follows. For each quasi-identifier, the size of all anonymity sets is obtained. Then, the frequency of each of these sizes was obtained, so that it becomes known how many anonymity sets are unique (that is, $k=1$), how many are twins, triples, etc. The number of anonymity sets that are unique divided by the total number of respondents is then the proportion of the sample that is sample unique ($Pr(SU)$), which can be used for comparison between different sets of quasi-identifiers. 

One disadvantage of this method is that the estimated month of birth, as determined earlier, can not be used properly. The reason for this is that in many cases, two possibilities exist for the month of birth. Attempting to construct anonymity sets calls for exact matches on the response pattern, but exact matches would in this case mean creating anonymity sets of respondents that have the same two possibilities for month of birth (assuming we remove all respondents with only one or zero possibilities). However, it is illogical to treat status of estimated month of birth (entirely missing, one possible value or two possible values) as a regular variable in an analysis of uniqueness, as it is not a realized value of on an attribute but rather it shows the uncertainty in the realized value of the attribute. One possibility is to select only the respondents which had one possible value as an estimated month of birth, but only 19.5\% of respondents were in this group, which would mean that most of the data would be removed by listwise deletion in this case. Estimated month of birth is thus not used in combination with this method. 



\vspace{15pt}
\noindent \textit{Number of matches}

Suppose an adversary has an individual's specific response pattern to be matched to the disclosed dataset. Then, this individual will (most likely) have the same values in the disclosed dataset as in the response pattern, which allows linking. However, the individual may also have neglected to enter information in the response pattern, causing a missing value. Thus, the adversary should match each of the variables in the quasi-identifier to the value that is in the response pattern, as well as to all missing values. 

A method of assessing re-identification risk was implemented using the scenario above as its basis. It involves going through the dataset respondent by respondent, calculating for each how many respondents in the dataset would match this value. The method to obtain the number of matches a respondent has for its own response pattern $j$ and on quasi-identifier $q$ can be defined as follows:

\begin{itemize}
\item[\textbf{Step 1}] For a variable in $q$, look up the corresponding value in $j$.
\item[\textbf{Step 2}] If the value in $j$ is not missing, remove all respondents from the dataset that do not have either the value specified in $j$ or a missing value on the variable in $q$.
\item[\textbf{Step 3}] If the value in $j$ is missing, do nothing.
\item[\textbf{Step 4}] Repeat step 1-3 for each remaining variable in $q$.
\item[\textbf{Step 5}] The number of respondents remaining in the dataset is then the number of matches, $n_{match}$, which has a minimum of 1, as the matches always include the respondent itself. 
\end{itemize}

As explained before, month of birth was estimated in the disclosed dataset. This caused an issue in addressing which value should be used for the corresponding value in $j$ in step 1 of the method. Whether another pattern matches the current row depends on this estimated month of birth. However, the method simulates the situation where an adversary tries to match any respondent in a linking dataset to a respondent in the disclosed dataset. In a linking dataset, it is expected that, if we are to use month of birth in the quasi-identifier, month of birth is known rather than estimated. Thus, in this matching method, estimated month of birth was treated as follows. If it was missing entirely, nothing was done (as in step 3). If there was one value as an estimated month of birth, this value was used as the value in $j$ corresponding to the variable month of birth in $q$. If there were two possible values, the first of the two was (arbitrarily) used as the corresponding value in $j$. 


The result of this method is a measure of uniqueness that is close to the actual scenario of an adversary attempting to re-identify a person which is known to be in the disclosed dataset. The measure denotes, for a specific quasi-identifier and for each respondent $r$, the amount of matches between the response patterns of the respondent $r$ and all other respondents. The number of matches is distinct from the size of the anonymity set $k$. 

As an example of this, consider Table \ref{tab:matchexample}, where $k$ denotes the anonymity set size under listwise deletion, and $n_{match}$ denotes the measure as described above. For $k$-anonymity, all matches must be exact matches. For $n_{match}$, values are allowed to match missing values in addition to exact matches. Because of this, $ n_{match} \geq k$ holds within the same dataset. 

\begin{table}
\begin{small}

\begin{center}
\renewcommand{\arraystretch}{1.2}
\begin{tabular}{r c|c c}
\\[-1.8ex]\hline 
\hline \\[-1.8ex] 
Age 		& Gender 	& $k$ & $n_{match}$\\ 
\hline
40			& Male		& 1 & 3 \\
36			& Female	& 1 & 1 \\
.			& Male		& - & 4 \\
40			& .			& - & 2 \\
23			& Male		& 2 & 3 \\
23			& Male		& 2 & 3 \\
\hline \vspace{-1cm}
\end{tabular}
\end{center}

\caption{Example of the difference between the measures $k$ and the proposed $n_{match}$. \label{tab:matchexample}}
\end{small}
\end{table}

The $n_{match}$ method is a novel way to assess re-identification risk which is geared towards data that contains missing values as well as missing questionnaires for a part of respondents. It has several advantages over the listwise deletion sample uniqueness method. First, respondents which have missing data can be maintained, which allows for better estimates of re-identification risk.  Second, the method more closely resembles the scenario in which an adversary would attempt re-identification, because it takes a practical issue into account that the adversary will run into. An additional advantage is that in situations where no missing data exists, the method is equal to the listwise deletion sample uniqueness method. 

In practice, this means that this method is more optimistic regarding anonymity, yet also more realistic under common scenarios for survey research specifically. It takes into account the fact that missing data strongly diminishes the usefulness of variables in the quasi-identifier, as they provide additional uncertainty. 

\subsection*{Results}

In this section, the previously described two methods of assessing risk of re-identification through linking datasets will be applied to the LISS data. 

\begin{center}
\begin{table}[!htbp] 
\begin{adjustwidth}{-2.6cm}{2cm}

\begin{small}
\begin{tabular}{@{\extracolsep{0pt}} lcccccccc} 
\\[-1.8ex]\hline 
\hline \\[-1.8ex] 
Quasi-Identifier & $n_{deleted}$ & n & $k = 1$ & $k \leq 5$ & $k \leq 10$& $Pr(SU)_{new}$ & $Pr(SU)_{full}$ & $\theta$ \\ 
\hline \\[-1.8ex] 
YoB 												&$0$ & $10,997$ & $3$ & $5$ & $6$ & $0.0003$ & $0.0003$ & $0.0010$ \\ 
YoB + Gender 										&$0$ & $10,997$ & $4$ & $9$ & $18$ & $0.0004$ & $0.0004$ & $0.0007$ \\ 
Children 											&$0$ & $10,997$ & $1$ & $1$ & $1$ & $0.0001$ & $0.0001$ & $-$ \\ 
Income 												&$1,783$ & $9,214$ & $856$ & $1,431$ & $1,506$ & $0.0929$ & $0.0778$ & $0.0007$ \\ 
Religion 											&$8,683$ & $2,314$ & $0$ & $1$ & $3$ & $0$ & $0$ & $-$ \\ 
Profession 											&$5,391$ & $5,606$ & $0$ & $0$ & $0$ & $0$ & $0$ & $-$ \\ 
Social Media 										&$5,802$ & $5,195$ & $0$ & $0$ & $0$ & $0$ & $0$ & $-$ \\ 
Language 											&$4,848$ & $6,149$ & $7$ & $19$ & $23$ & $0.0011$ & $0.0006$ & $0.0002$ \\ 
\hline \\[-1.8ex]

YoB + Gender + Occupation							&$0$ & $10,997$ & $269$ & $564$ & $667$ & $0.0245$ & $0.0245$ & $0.0007$ \\ 
YoB + Gender + Income cat.							&$406$ & $10,591$ & $262$ & $609$ & $785$ & $0.0247$ & $0.0238$ & $0.0007$ \\ 
YoB + Gender + Income 								&$1,783$ & $9,214$ & $4,616$ & $5,813$ & $5,860$ & $0.5010$ & $0.4198$ & $0.0017$ \\ 
YoB + Gender + Origin 								&$2,264$ & $8,733$ & $107$ & $360$ & $410$ & $0.0123$ & $0.0097$ & $0.0003$ \\ 
YoB + Gender + Children 							&$0$ & $10,997$ & $99$ & $216$ & $289$ & $0.0090$ & $0.0090$ & $0.0006$ \\ 
YoB + Gender + Religion								&$8,683$ & $2,314$ & $271$ & $571$ & $657$ & $0.1171$ & $0.0246$ & $0.0001$ \\ 
YoB + Gender + Language								&$4,848$ & $6,149$ & $409$ & $597$ & $610$ & $0.0665$ & $0.0372$ & $0.0006$ \\ 
YoB + Gender + Educ. Field 							&$5,131$ & $5,866$ & $1,165$ & $1,825$ & $1,982$ & $0.1986$ & $0.1059$ & $0.0006$ \\ 
YoB + Gender + Profession 							&$5,391$ & $5,606$ & $233$ & $689$ & $905$ & $0.0416$ & $0.0212$ & $0.0003$ \\ 
YoB + Gender + Social Media							&$5,802$ & $5,195$ & $153$ & $519$ & $654$ & $0.0295$ & $0.0139$ & $0.0002$ \\ 
\hline \\[-1.8ex]
YoB+Gender+Occ.+Prof.+Educ. Field 					&$5,499$ & $5,498$ & $3,353$ & $4,044$ & $4,098$ & $0.6099$ & $0.3049$ & $0.0012$ \\ 
YoB+Gender+Occ.+Prof.+Educ. Field+Income cat.		&$5,691$ & $5,306$ & $4,258$ & $4,697$ & $4,700$ & $0.8025$ & $0.3872$ & $0.0020$ \\ 
YoB+Gender+Origin+Children+Soc. Med. 				&$5,929$ & $5,068$ & $1,209$ & $1,905$ & $1,989$ & $0.2386$ & $0.1099$ & $0.0005$ \\ 
YoB+Gender+Origin+Children+Soc. Med.+Lang. 			&$6,141$ & $4,856$ & $1,541$ & $2,165$ & $2,239$ & $0.3173$ & $0.1401$ & $0.0006$ \\ 

\hline \\[-1.8ex] 
\normalsize 
\end{tabular} 
\end{small}
\vspace{-15pt}
  \caption{Number of respondents per anonymity set size $k$, for various quasi-identifiers} 
  \label{tab:listwise} 
\end{adjustwidth}
\end{table} 

\end{center}

\noindent \textit{Listwise deletion}

In Table \ref{tab:listwise} results are shown for several sets of quasi-identifiers that were tested. First, the quasi-identifiers are given. \textit{YoB} refers to year of birth, \textit{Children} to the amount of living-at-home children, \textit{Income} to the net monthly income. \textit{Profession (Prof.)} refers to current type of profession in nine possible categories. 

\textit{Religion}, \textit{social media}, \textit{language} and \textit{educational field} were all constructed from multiple questions in the disclosed dataset. This was done because in a linking dataset it would be expected that all elements from the set of questions could be found if any one of them was found. \textit{Religion} consists of the questions whether a respondent follows a religion and if so, which. \textit{Social Media (Soc. Med.)} refers to whether a respondent has an online profile on three possible social networks: Facebook, LinkedIn and Hyves (a Dutch social network site). This was grouped as the existence of a person's profile on these sites is often publicly visible online, even if the profile itself is not public. \textit{Language (Lang.)} consists of a list of nine questions asking whether a respondent grew up speaking a specific language. \textit{Educational field (Educ. Field)} consists of 17 dichotomous questions where respondents reported the field in which they had completed their highest level of education. 

$n_{deleted}$ here refers to the amount of rows that were deleted by the listwise deletion procedure, while $n$ is the amount of respondents that are left over and used in the analysis. Their sum is always equal to 10997, which is the amount of respondents in the working dataset before deletion. 

Then, the frequency of $k=1$, or unique respondents, is shown. This denotes the frequency of anonymity sets that contain only a single respondent. The columns $k \leq 5$ and $k \leq 10$ give the number of respondents that fall into an anonymity set of size smaller than 5 and 10 respectively. From this, we can determine how common it is for respondents to be indistinguishable from only a few people. 

$Pr(SU)$ in general denotes the proportion of the sample that is sample unique. Here, $Pr(SU)_{new}$ shows the proportion of respondents that is sample unique after listwise deletion (that is, the number given under $k=1$ divided by the number given under $n$), while $Pr(SU)_{full}$ gives the proportion of the full sample (of 10997 respondents) that was shown to be unique with this method (that is, the number given under $k=1$ divided by 10997). As more variables are added to the quasi-identifier, $Pr(SU)_{new}$ will usually increase, while $Pr(SU)_{full}$ may increase or remain stable but can also drop if the added variable contains missing values. Then, potentially unique respondents may be removed by listwise deletion and the proportion of the full sample that is found to be unique may drop. In general, $Pr(SU)_{full}$ is the most interesting as it is tied to re-identification risk in the complete dataset, while $Pr(SU)_{new}$ gives an idea of the proportion of respondents that would be unique if the disclosed dataset did not contain missings. 

Finally, the $\theta$ as given by \cite{skinner2002measure} was calculated and is also given. For the situations with very few unique respondents it was unreliable, and thus removed.

In the table, several possible quasi-identifiers were used to see which provide most uniqueness. They were chosen to show properties of several likely quasi-identifiers in practice. For example, the final four are chosen to represent a possible linking dataset that contains information on a either respondent's professional or personal life. 

In general, several properties of variables can increase the amount of uniqueness they add. First, variables containing a large amount of variables, or, even better, continuous variables, provide more uniqueness. Second, evenly distributed variables (e.g. 25\%/25\%/25\%/25\% instead of 85\%/5\%/5\%/5\%) provide more uniqueness. Respondents with rare values may of course become unique, but in combination with other variables, an even distribution between the categories excludes more matches for most respondents, causing the variable to add more uniqueness. Thirdly, variables that are independent of other variables in the quasi-identifier provide more uniqueness, because two correlated variables provide less information than two independent variables. 

This can also be seen in Table \ref{tab:listwise}. Most categorical variables provide no uniqueness by themselves, so these rows are not shown. Income provides a large amount of unique respondents by itself, as it is a continuous variable. However, it is fairly unlikely that income is matched exactly between two datasets: first, respondents may provide an estimate instead of an exact number, second, it is subject to change, and third, it may be calculated in different ways. 

Year of birth by itself grants three unique cases, which are most likely elderly respondents, as for them the distribution should be fairly sparse. As most quasi-identifiers consisting of a single variable provide few unique respondents and are also unlikely, year of birth and gender are then combined with several other variables.

Professional information was quite useful in quasi-identifiers. The quasi-identifier consisting of year of birth, gender, occupation, profession, educational field and income in categories showed the highest proportion of unique values in the sample after listwise deletion, 80.25\%. Moreover, it also showed a very high proportion of unique respondents from the full sample: 38.7\% of the original sample of 10997 could be uniquely identified by this quasi-identifier. A similar pattern can be seen looking at $\theta$. If the response pattern is known of a respondent and an adversary finds a match between a respondent and a person from the population on this quasi-identifier, not knowing whether this person is in the LISS panel, there is a probability of .2\% that the match will be correct. 

Personal background information was slightly less useful, with the quasi-identifier containing year of birth, gender, origin, number of living-at-home children, social media presence and languages uniquely identified 14\% of the original sample. 

\begin{table}[!htbp] \centering 
\begin{adjustwidth}{-2.2cm}{2cm}

\begin{tabular}{@{\extracolsep{2pt}} lccccc} 
\\[-1.8ex]\hline 
\hline \\[-1.8ex] 
Quasi-identifier & $n_{match} = 1$ & $n_{match} \leq 5$ & $n_{match} \leq 10$& Pr(SU) & $\theta$ \\ 
\hline \\[-1.8ex] 
YoB 												&$3$ & $8$ & $15$ & $0.0003$ & $0.0005$ \\              
YoB + Gender 										&$4$ & $20$ & $85$ & $0.0004$ & $0.0003$ \\             
Children 											&$1$ & $1$ & $1$ & $0.0001$ & $-$ \\                    
Income 												&$0$ & $0$ & $0$ & $0$ & $-$ \\                         
Religion 											&$0$ & $0$ & $0$ & $0$ & $-$ \\                         
Profession 											&$0$ & $0$ & $0$ & $0$ & $-$ \\                         
Social Media 										&$0$ & $0$ & $0$ & $0$ & $-$ \\                         
Language 											&$0$ & $0$ & $0$ & $0$ & $-$ \\                         
\hline \\[-1.8ex]
YoB + Gender + Occupation							&$269$ & $1,151$ & $1,944$ & $0.024$ & $0.0004$ \\      
YoB + Gender + Income cat.							&$66$ & $618$ & $1,924$ & $0.006$ & $0.0002$ \\         
YoB + Gender + Income 								&$49$ & $307$ & $1,393$ & $0.004$ & $0.0003$ \\         
YoB + Gender + Origin 								&$5$ & $40$ & $146$ & $0.0005$ & $0.0002$ \\            
YoB + Gender + Children 							&$99$ & $463$ & $1,042$ & $0.009$ & $0.0003$ \\         
YoB + Gender + Religion								&$4$ & $36$ & $119$ & $0.0004$ & $0.0003$ \\            
YoB + Gender + Language								&$4$ & $31$ & $106$ & $0.0004$ & $0.0003$ \\            
YoB + Gender + Educ. Field 							&$4$ & $38$ & $168$ & $0.0004$ & $0.0001$ \\            
YoB + Gender + Profession 							&$4$ & $32$ & $112$ & $0.0004$ & $0.0003$ \\            
YoB + Gender + Social Media							&$4$ & $20$ & $89$ & $0.0004$ & $0.0003$ \\             
\hline \\[-1.8ex]
YoB + MoB + Gender + Occupation						&$1,559$ & $5,377$ & $9,460$ & $0.142$ & $0.0005$ \\ 
YoB + MoB + Gender + Income cat.					&$1,789$ & $8,908$ & $10,611$ & $0.163$ & $0.0003$ \\
YoB + MoB + Gender + Income 						&$1,567$ & $8,194$ & $9,982$ & $0.142$ & $0.0002$ \\ 
YoB + MoB + Gender + Origin 						&$163$ & $2,508$ & $7,068$ & $0.015$ & $0.0001$ \\   
YoB + MoB + Gender + Children 						&$1,123$ & $6,553$ & $9,806$ & $0.102$ & $0.0002$ \\ 
YoB + MoB + Gender + Religion						&$139$ & $2,458$ & $7,199$ & $0.013$ & $0.0002$ \\   
YoB + MoB + Gender + Language						&$103$ & $1,849$ & $6,462$ & $0.009$ & $0.0002$ \\   
YoB + MoB + Gender + Educ. Field 					&$178$ & $3,265$ & $7,967$ & $0.016$ & $0.0001$ \\   
YoB + MoB + Gender + Profession 					&$131$ & $2,769$ & $7,548$ & $0.012$ & $0.0001$ \\   
YoB + MoB + Gender + Social Media					&$82$ & $2,117$ & $6,984$ & $0.008$ & $0.0001$ \\    
\hline \\[-1.8ex]
YoB+Gender+Occ.+Prof.+Educ. Field 				&$430$ & $1,600$ & $2,659$ & $0.039$ & $0.0004$ \\   
YoB+Gender+Occ.+Prof.+Educ. Field+Income cat.	&$1,369$ & $4,436$ & $6,663$ & $0.124$ & $0.0005$ \\ 
YoB+Gender+Origin+Children+Soc. Med. 			&$185$ & $893$ & $2,132$ & $0.017$ & $0.0004$ \\ 
YoB+Gender+Origin+Children+Soc. Med.+Lang. 		&$192$ & $959$ & $2,263$ & $0.018$ & $0.0004$ \\ 
\hline \\[-1.8ex]
YoB+MoB+Gender+Occ.+Prof.+Educ. Field 				&$2,131$ & $7,357$ & $10,239$ & $0.194$ & $0.0004$ \\ 
YoB+MoB+Gender+Occ.+Prof.+Educ. Field+Income cat.	&$5,146$ & $10,391$ & $10,873$ & $0.468$ & $0.001$ \\ 
YoB+MoB+Gender+Origin+Children+Soc. Med. 			&$1,823$ & $7,931$ & $10,339$ & $0.166$ & $0.0003$ \\ 
YoB+MoB+Gender+Origin+Children+Soc. Med.+Lang. 		&$1,912$ & $8,089$ & $10,398$ & $0.174$ & $0.0003$ \\ 
\hline \\[-1.8ex] 
\normalsize 
\end{tabular} 
\vspace{-15pt}
  \caption{Frequencies of $n_{match}$ for various quasi-identifiers} 
  \label{tab:nmatch} 
\end{adjustwidth}
\end{table} 

\vspace{15pt}

\noindent \textit{Number of matches}

As explained before, listwise deletion is not the only way to assess re-identification risk in a dataset linking scenario, and for survey panels the $n_{match}$ method may be have several advantages. In Table \ref{tab:nmatch} the results for this method are shown. No respondents were removed in with this method, so $n = 10997$ for each quasi-identifier. 

In this table, the column $n_{match} = 1$ denotes the amount of respondents that matched on the given quasi-identifier with only itself. This is analogous to a respondent having $k = 1$ in the sense that these values allow an adversary to determine a single respondent which is likely a correct match.  $n_{match} \leq 5$ and $n_{match} \leq 10$ refer to the amount of respondents that have less than, respectively, 5 and 10 matches. $Pr(SU)$ refers to the proportion of the population that has a single match, where $\theta$ is calculated as before, now using the number of matches. 

The methods differ in several regards. In listwise deletion, income was very useful as we would only use those respondents that were not missing, and knowing income would uniquely identify a large amount of respondents. Here, as any value of income is also matched to missing values, it does not provide unique values by itself. However, it does exclude values that differ from a respondents value (if it is non-missing), so it may increase the frequency of $n_{match} = 1$, $n_{match} \leq 5$, and $n_{match} \leq 10$ in combination with other variables. 

In general, the frequency of respondents with $n_{match} = 1$ is lower than the frequency of respondents with $k = 1$. This is caused by the fact that any value in the $n_{match}$ method is also matched to missing values, while for listwise deletion these would have been deleted and as such not part of that respondents' anonymity set. In general, then, lower values for $Pr(SU)$ are reported for the $n_{match}$ method compared to listwise deletion, with the exception of the quasi-identifiers for which a very large part of the sample had to be omitted in the listwise deletion method.

Here, estimated month of birth (MoB) was also used, as the $n_{match}$ method allows for use of variables that contain uncertainty. Estimated month of birth contributed a large amount of uniqueness. For example, the quasi-identifier consisting of year of birth, gender and occupation gives 269 respondents that only match themselves, 2.4\% of the sample. Adding estimated month of birth gives 1559 respondents that only match themselves, 14.2\% of the sample. This large increase can be attributed to the fact that month of birth has 12 categories which are evenly distributed and not correlated with other variables. Because of this, it provides a very strong contribution to proportion of respondents that only match themselves. 

As can be seen, the $\theta$ for all quasi-identifiers is fairly low. Interpreting it as the probability that a unique match is correct, this probability is found to be at most .1\%. This probability deals with the scenario where an adversary would attempt to re-identify a respondent on the basis of a match between a person drawn from the population and a respondent in the disclosed dataset. Because it is not known in that situation whether the person from the population is in the disclosed dataset, it remains very unlikely that this is in fact the case. This illustrates the fact that knowledge that the person that is being re-identified is in the disclosed dataset helps re-identification efforts tremendously. While it may be possible to identify respondents by linking them with another dataset, it requires very strong quasi-identifiers that have been measured with a large amount of certainty.

\section{Concluding remarks \label{conclusion}}

Re-identification efforts in survey panels were generally shown to be slightly more difficult than in most other datasets that are used in disclosure control literature, which are often medical or register data. For survey data usually a larger amount of uncertainty exists in the data, and a larger amount of missings, due to both respondents not answering questions as well as questionnaires being administered to subsamples. The number of matches method was shown to be more realistic for survey data, as it correctly shows the result of missing values on re-identification efforts. It shows much smaller proportions of unique respondents and thus more difficult re-identification. Regardless, with enough information, re-identification may be possible. 

When determining what should be done to enforce anonymity in a survey panel, a trade-off must then be made between loss of utility (e.g. analyses that can be done on the data may decline in quantity and quality) and certainty of anonymity. The potential loss 
of proposed steps to be taken to maintain anonymity may be estimated while also estimating the effect of our measures on anonymity. In general, precautions that have little to no effect on utility while hindering re-identification efforts should always be taken, while for other steps, such as aggregation, censoring variables or enforcing a cut-off point, it must be investigated whether the gain in anonymity is worth the loss of utility. 

\section{Acknowledgments}

We would like to thank Annette Scherpenzeel for helpful discussions and comments, as well as Eric Balster and other employees at CentERdata for providing support and access to the LISS-panel.

\bibliographystyle{plain}
\bibliography{EA}

\end{document}